\documentclass[aps,prl,reprint,groupedaddress,nofootinbib]{revtex4-1}
\pdfoutput=1 
\usepackage{graphicx}
\usepackage{color}
\usepackage{rotating}
\usepackage{amsmath}
\usepackage{amssymb}
\usepackage[dvipsnames]{xcolor}

\newcommand{\beq}{\begin{equation}}
\newcommand{\eeq}{\end{equation}}
\newcommand{\bea}{\begin{eqnarray}}
\newcommand{\eea}{\end{eqnarray}}

\newcommand{\MeV}{\, \textrm{MeV}}
\newcommand{\GeV}{\, \textrm{GeV}}

\allowdisplaybreaks

\begin{document}
\title{Lattice-based equation of state at finite baryon number, electric charge and strangeness chemical potentials}

\author{J. Noronha-Hostler$^{a,b}$, P. Parotto$^{c,d}$, C. Ratti$^d$, J. M. Stafford$^d$}
\affiliation{\small{\it $^a$ Department of Physics and Astronomy, Rutgers University, Piscataway, NJ USA 08854}}
\affiliation{\small{\it $^b$Department of Physics, University of Illinois at Urbana-Champaign, Urbana, IL 61801, USA}}
\affiliation{\small{\it $^c$ University of Wuppertal, Department of Physics, Wuppertal D-42097, Germany}}
\affiliation{\small{\it $^d$ Department of Physics, University of Houston, Houston, TX, USA  77204}}

\date{\today}
\
\begin{abstract}
We construct an equation of state for Quantum Chromodynamics (QCD) at finite temperature and chemical potentials for baryon number $B$, electric charge $Q$ and strangeness $S$. We use the Taylor expansion method, up to the fourth power for the chemical potentials. This requires the knowledge of all diagonal and non-diagonal $BQS$ correlators up to fourth order: these results recently became available from lattice QCD simulations, albeit only at a finite lattice spacing $N_t=12$. We smoothly merge these results to the Hadron Resonance Gas (HRG) model, to be able to reach temperatures as low as 30 MeV; in the high temperature regime, we impose a smooth approach to the Stefan-Boltzmann limit. We provide a parameterization for each one of these $BQS$ correlators as functions of the temperature. We then calculate pressure, energy density, entropy density, baryonic, strangeness, electric charge densities and compare the two cases of strangeness neutrality and $\mu_S=\mu_Q=0$. Finally, we calculate the isentropic trajectories and the speed of sound, and compare them in the two cases. Our equation of state can be readily used as an input of hydrodynamical simulations of matter created at the Relativistic Heavy Ion Collider (RHIC).
\end{abstract}
\pacs{}

\maketitle

\section{Introduction}

Relativistic heavy ion collisions have successfully recreated the Quark Gluon Plasma (QGP) in the laboratory at the Relativistic Heavy Ion Collider (RHIC) at Brookhaven National Laboratory and the Large Hadron Collider (LHC) at CERN.  At low baryon densities, the transition from the hadron gas phase where quarks and gluons are confined within hadrons into a deconfined state where quark and gluons are the main degrees of freedom is a smooth cross-over \cite{Aoki:2006we,Borsanyi:2010bp,Bazavov:2011nk}. At larger baryon densities, the phase transition is expected to become stronger, eventually turning into first-order.  If this is the case, there has to be a critical point on the QCD phase diagram \cite{Stephanov:1998dy,Halasz:1998qr,Stephanov:1999zu,McLerran:2007qj,Critelli:2017oub}. The search for the QCD critical point is the focus of the second Beam Energy Scan (BES II) at RHIC, running in 2019 and 2020.

The Quark Gluon Plasma acts as a nearly perfect fluid and as such can be well-described by event-by-event relativistic viscous hydrodynamical models. 
The hydrodynamical description of the fireball has proved to be very successful in describing the experimental data \cite{Luzum:2008cw,Gale:2012rq,Gardim:2012yp,Song:2013qma,Niemi:2015voa,Niemi:2015qia,Noronha-Hostler:2015uye,Bernhard:2016tnd,Alba:2017hhe,McDonald:2016vlt,Giacalone:2017dud}. In order to close the hydrodynamical equations, an Equation of State (EoS) is required, which is based on first principle Lattice QCD calculations. Recently, a Bayesian analysis \cite{Pratt:2015zsa} has provided an important validation of the lattice QCD equation of state.   This framework, based on a comparison of data
from the LHC to theoretical models, has applied state-of-the-art statistical
techniques to the combined analysis of a large number of observables while varying the model parameters. The posterior distribution over possible equations of states turned out to be consistent with results from lattice QCD simulations.  Additionally, the correct description of the QCD equation of state is needed because differences in the equation can affect the extraction of transport coefficients \cite{Alba:2017hhe}. Thus, a lattice-based QCD equation of state is a fundamental ingredient in the description of the state of matter created in a heavy-ion collision. The precise lattice QCD results for several thermodynamic quantities can thus be used in support of the heavy ion experimental program \cite{Ratti:2018ksb}.

The EoS of QCD at zero baryonic density is known with high precision from first principles since a few years \cite{Borsanyi:2010cj,Borsanyi:2013bia,Bazavov:2014pvz}. The calculation of the equation of state at finite chemical potential is hindered by the sign problem. Nevertheless, the thermodynamic quantities can be expanded as a Taylor series in powers of $\mu_B/T$, for which the coefficients $\chi_n$ can be simulated on the lattice at $\mu_B=0$. From these Taylor coefficients a variety of Lattice QCD based equations of state have been reconstructed \cite{Monnai:2012jc,Parotto:2018pwx,Vovchenko:2018zgt}  and later used within relativistic hydrodynamics \cite{Monnai:2012jc,Karpenko:2015hea,Batyuk:2016qmb,Monnai:2016kud,Denicol:2018wdp}. 

However, baryon number is not the only conserved charge in a heavy ion collision: strangeness and electric charge are also relevant quantum numbers. In fact, many questions remain regarding a possible separate freeze-out temperature for strange hadrons \cite{Bellwied:2013cta,Noronha-Hostler:2016rpd,Ratti:2018vxc,Bluhm:2018aei} and separations of electric charge due to a possible chiral magnetic effect \cite{Fukushima:2008xe}, so many interesting questions need to be answered, that go beyond just baryon charge conservation.  
At the LHC, where the baryonic chemical potential $\mu_B$ is basically vanishing, the chemical potentials for strangeness $\mu_S$ and electric charge $\mu_Q$ are also zero. At RHIC however, as the baryonic density increases, the other two chemical potentials have finite values as well. Until now, the equation of state of QCD has only been extrapolated to finite $\mu_B$, either by keeping $\mu_S=\mu_Q=0$, or along a specific trajectory in the four-dimensional parameter space, namely imposing that the strangeness density $\langle n_S\rangle=0$ and that the electric charge density $\langle n_Q\rangle=0.4\langle n_B\rangle$ to match the experimental situation.

After the early results for $\chi_2,~\chi_4$ and $\chi_6$ \cite{Allton:2005gk}, a continuum extrapolation for $\chi_2$ was published in Ref. \cite{Borsanyi:2012cr}; in Ref. \cite{Hegde:2014sta} $\chi_4$ was shown, but only at finite lattice spacing. The continuum limit for $\chi_6$ was published for the first time in \cite{Gunther:2016vcp} in the case of strangeness neutrality, and later in \cite{Bazavov:2017dus} for both cases. In \cite{DElia:2016jqh}, a first determination of $\chi_8$ at two values of the temperature and $N_t=8$ was presented. Finally, in Ref. \cite{Borsanyi:2018grb} a determination of $\chi_8$ was presented for the first time as a function of the temperature, at $N_t=12$, keeping $\mu_S=\mu_Q=0$. Recently, the effect of introducing a critical point in the equation of state of QCD has also been tested \cite{Parotto:2018pwx}.

However, a Taylor expansion of the equation of state, along a direction which satisfies the strangeness-neutrality condition is not enough for the hydrodynamics approach, since the fluid cells have local fluctuations in strangeness density.   Additionally, there is a complicated interplay between transport coefficients when $B,~Q,~S$ are considered \cite{Greif:2017byw} that cannot be neglected at large baryon densities.  For these reasons, an EoS fully expanded as a Taylor series in powers of $\mu_B/T,~\mu_S/T,~\mu_Q/T$ is needed as an input of hydrodynamic simulations of the matter created at RHIC. In order to perform such an expansion, all of the diagonal and non-diagonal susceptibilities of these three conserved charges are needed from lattice QCD up to the chosen power. In this work, we perform the Taylor expansion of to total power four in the chemical potentials. These results recently became available \cite{Borsanyi:2018grb}, on $N_t=12$ lattices.

Alternative approaches to the Taylor series expansion have been suggested in \cite{Vovchenko:2018eod,Motornenko:2019arp} and \cite{Vovchenko:2016rkn,Vovchenko:2017cbu}, which have been shown to match well to lattice QCD data for the Fourier harmonics \cite{Vovchenko:2017xad} at imaginary chemical potential. These Fourier harmonics appear to be important to distinguish baryon interactions within a hadron resonance gas (see also \cite{Huovinen:2017ogf}), specifically for the thermodynamic regime above $T > 150 \MeV$. We note that here we use lattice QCD data entirely in this regime (our hadron resonance gas model is only to constrain low temperatures below $T \lesssim135 \MeV$ where no lattice QCD results are available). However, due to the Taylor expansion our approach is limited to chemical potentials $\mu_B \lesssim (2-2.5) \, T$. To fully reproduce the Fourier harmonics we would need to reach $\mu_B \lesssim \pi T$, for which higher order coefficients in the Taylor series would need to be included.

In this manuscript, we construct an equation of state for QCD at finite $T$, $\mu_B$, $\mu_S$, $\mu_Q$. We build the pressure as a Taylor series of the three chemical potentials, with coefficients taken from lattice simulations \cite{Borsanyi:2018grb}. At low temperatures, we perform a smooth merging between the lattice and the Hadron Resonance Gas model results \cite{Alba:2017mqu} and ensure continuity of higher order derivatives. At high temperatures, we impose a smooth approach to the Stefan-Boltzmann limit. We parameterize each one of these coefficients as a ratio of polynomials. From this we obtain the pressure  and can  then calculate all other quantities from thermodynamic relationships \footnote{Right before publishing this manuscript, we became aware of Ref. \cite{Monnai:2019hkn} which constructs a similar equation of state as the one presented here. One major difference is that we match lattice QCD susceptibilities with the hadron resonance gas model before reconstructing the equation of state whereas in \cite{Monnai:2019hkn} the matching with the HRG model is performed for the Taylor-reconstructed pressure. }.

\section{Methodology and results}
The Taylor series of the pressure in terms of the three conserved charge chemical potentials can be written as
\bea
\frac{p(T,\mu_B,\mu_Q,\mu_S)}{T^4}&=&\sum_{i,j,k}\frac{1}{i!j!k!}\chi_{ijk}^{BQS}
\left(\frac{\mu_B}{T}\right)^i\left(\frac{\mu_Q}{T}\right)^j\left(\frac{\mu_S}{T}\right)^k.
\label{p}
\nonumber\\
\eea
We limit our calculation to $i+j+k\leq4$. The coefficients
\bea
\chi_{ijk}^{BQS}=\left.\frac{\partial^{i+j+k}(p/T^4)}{\partial(\frac{\mu_B}{T})^i\partial(\frac{\mu_Q}{T})^j\partial(\frac{\mu_S}{T})^k}\right|_{\mu_B,\mu_Q,
\mu_S=0}
\eea
have recently been published from lattice QCD simulations on $48^3\times12$ lattices \cite{Borsanyi:2018grb} in the temperature range (135~MeV)$<T<(220$ MeV).
Since this is not enough to cover the hydrodynamical evolution of the system, we smoothly merge each coefficient at low temperature with the Hadron Resonance Gas model result, while at high temperature we calculate the Stefan-Boltzman limit for each one of them and assume that their value at $T=800$ MeV is $\sim10\%$ away from the respective Stefan-Boltzmann limit. 
To simplify the notation, whenever $i$, $j$, $k$ are zero, we only write the non-zero indices and only the corresponding conserved charges: for example, $\chi_{200}^{BQS}$ becomes $\chi_2^B$, $\chi_{301}^{BQS}$ becomes $\chi_{31}^{BS}$ and so on.
In order to provide a smooth pressure which can be easily derived to obtain the other thermodynamic quantities, we parameterize each coefficient by means of a ratio of up-to-ninth order polynomials in the inverse temperature:
\begin{widetext} 
\bea
\chi_{ijk}^{BQS}(T) &=& \frac{a^i_0 + a^i_1/t + a^i_2/t^2 + a^i_3/t^3 + a^i_4/t^4 + a^i_5/t^5+ a^i_6/t^6+ a^i_7/t^7+ a^i_8/t^8+ a^i_9/t^9}{b^i_0 + b^i_1/t + b^i_2/t^2 + b^i_3/t^3 + b^i_4/t^4 + b^i_5/t^5+b^i_6/t^6+b^i_7/t^7+b^i_8/t^8+b^i_9/t^9}+ c_0.
\nonumber
\label{param}
\eea
\end{widetext} 
Only $\chi_2^B$ requires a different parameterization:
\begin{equation} \label{eq:param_chi2B}
\chi^2(T) = e^{-h_1/t^\prime - h_2/{t^\prime}^2} \cdot f_3 \cdot (1 + \tanh(f_4 t^\prime + f_5))
\end{equation}
In both equations above, $t = T/154 \, \text{MeV}$, $t^\prime = T/200 \, \text{MeV}$ \cite{Borsanyi:2011sw}.
The values of the parameters for each coefficient are given in the appendix, together with the respective Stefan-Boltzmann limits. 
\begin{figure*}[t]
\centering
\includegraphics[width=\linewidth]{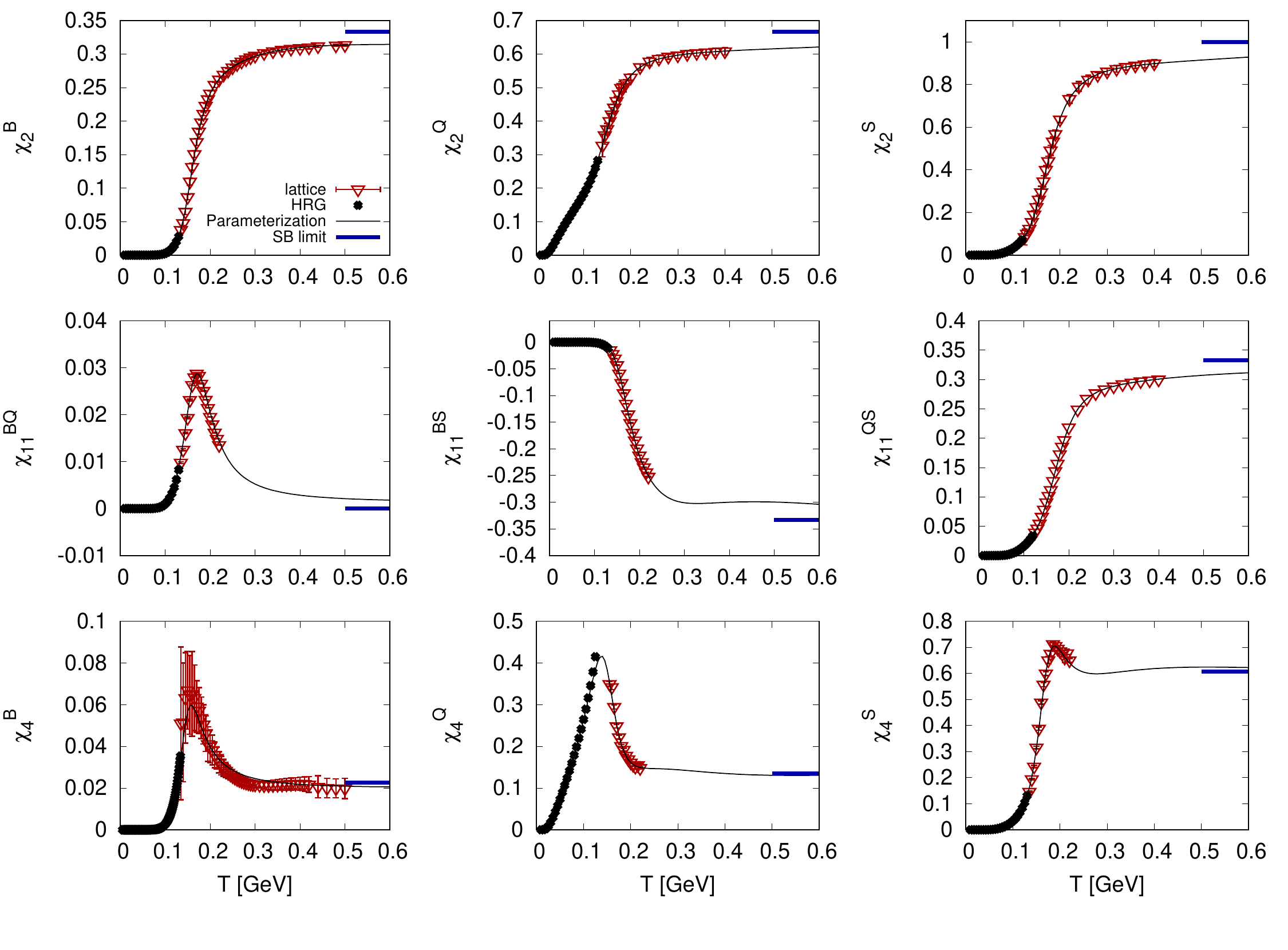} 
\caption{From left to right, top to bottom: expansion coefficients $\chi_2^B,~\chi_2^Q,\chi_2^S,~\chi_{11}^{BQ},~\chi_{11}^{BS},\chi_{11}^{QS},~\chi_4^B,~\chi_4^Q,\chi_4^S$ as functions of the temperature. In each panel, the black dots are the HRG model results, the red triangles correspond to the lattice QCD results and the thicker blue line on the right indicates the Stefan-Boltzmann limit. The thin solid, black curve shows our parameterization of the data.}
\label{fig1}
\end{figure*}
\begin{figure*}[t]
\centering
\includegraphics[width=\linewidth]{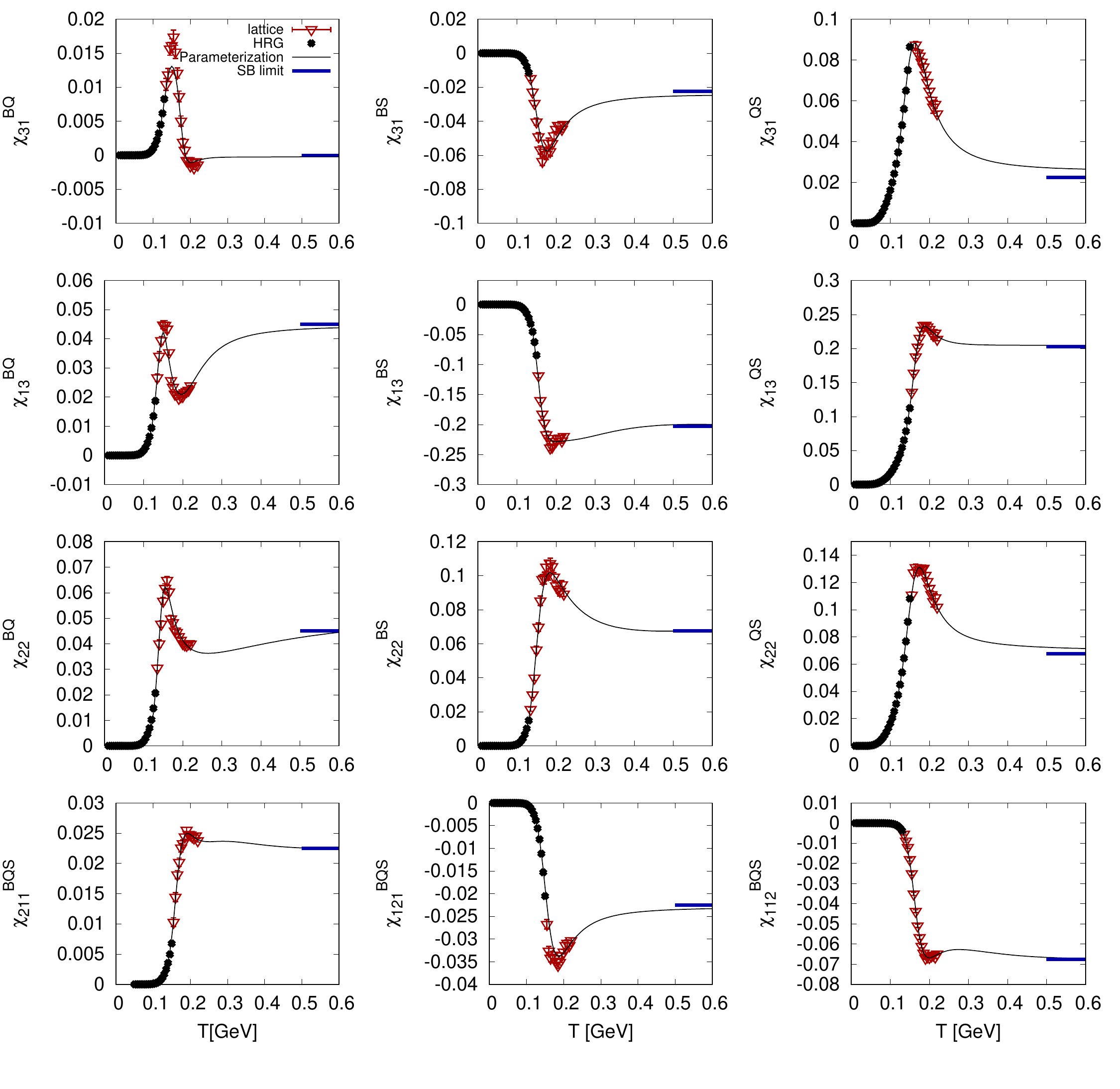} 
\caption{From left to right, top to bottom: expansion coefficients $\chi_{31}^{BQ},~\chi_{31}^{BS},~\chi_{31}^{QS},~\chi_{13}^{BQ},~\chi_{13}^{BS},~\chi_{13}^{QS},~\chi_{22}^{BQ},~\chi_{22}^{BS},~\chi_{22}^{QS},\chi_{211}^{BQS},~\chi_{121}^{BQS},$ $~\chi_{112}^{BQS},$ as functions of the temperature. In each panel, the black dots are the HRG model results, the red triangles correspond to the lattice QCD results and the thicker blue line on the right indicates the Stefan-Boltzmann limit. The thin solid, black curve shows our parameterization of the data.}
\label{fig2}
\end{figure*}
Figures \ref{fig1} and \ref{fig2} show all of the Taylor expansion coefficients as functions of the temperature. The black dots are the HRG model results, the red triangles correspond to the lattice QCD results and the thick blue line indicates the Stefan-Boltzmann limit.

Making use of this parameterization, we construct the pressure from Eq. (\ref{p}). The other thermodynamic quantities are then derived from the pressure as follows:
\bea
\frac{s}{T^3}&=&\left.\frac{1}{T^3}\frac{\partial p}{\partial T}\right|_{\mu_i},~~~~~~\frac{\epsilon}{T^4}=\frac{s}{T^3}-\frac{p}{T^4}+\sum_{i}\frac{\mu_i}{T}\frac{n_i}{T^3}
\nonumber\\
\!\!\!\!\!\!\frac{n_i}{T^3}&=&\frac{1}{T^3}\left. \frac{\partial p}{\partial \mu_i} \right|_{T,\mu_j},~~~c_s^2=\left.\frac{\partial p}{\partial \epsilon}\right|_{n_i}+\sum_{i}\frac{n_i}{\epsilon+p}\left.\frac{\partial p}{\partial n_i}\right|_{\epsilon, n_j}.
\eea
Everywhere in the above equation, $i\neq j$ is intended. 

\begin{figure*}[h] 
\centering
\includegraphics[width=\linewidth]{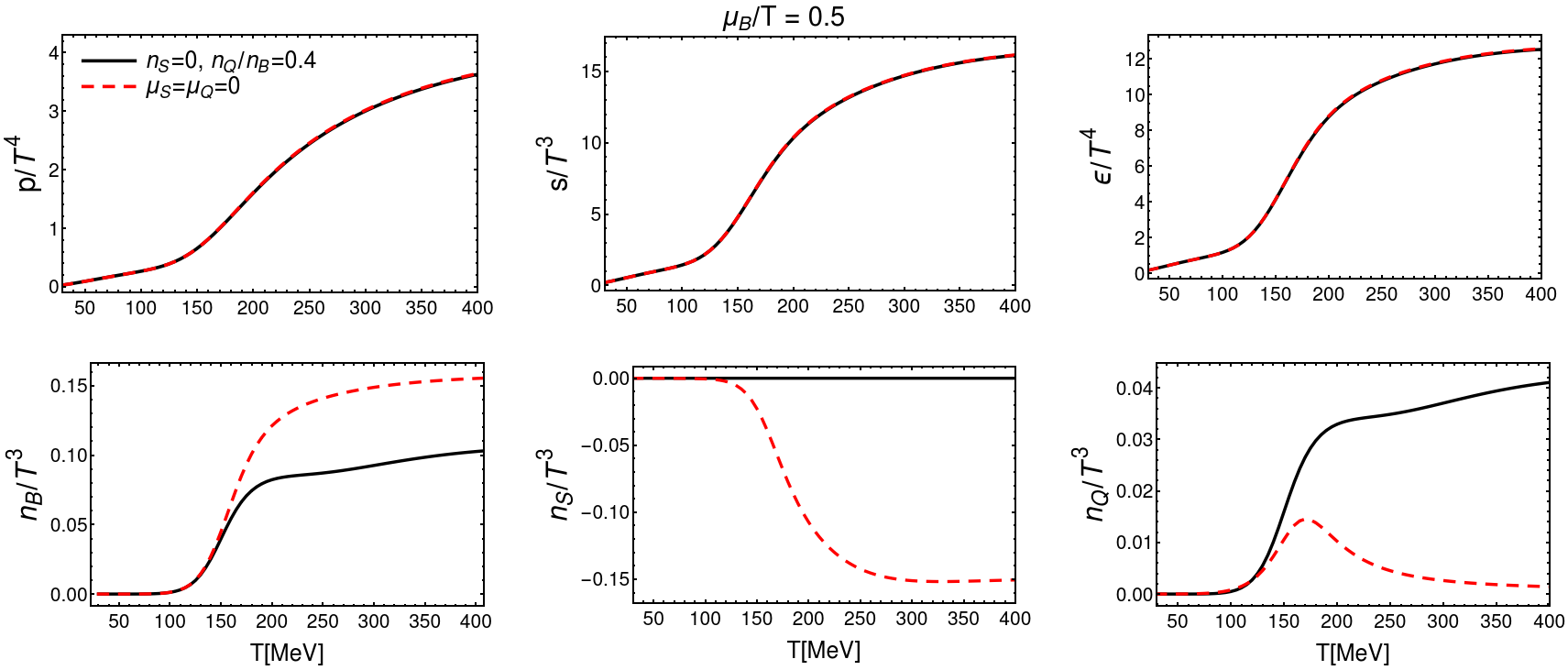} 

\includegraphics[width=\linewidth]{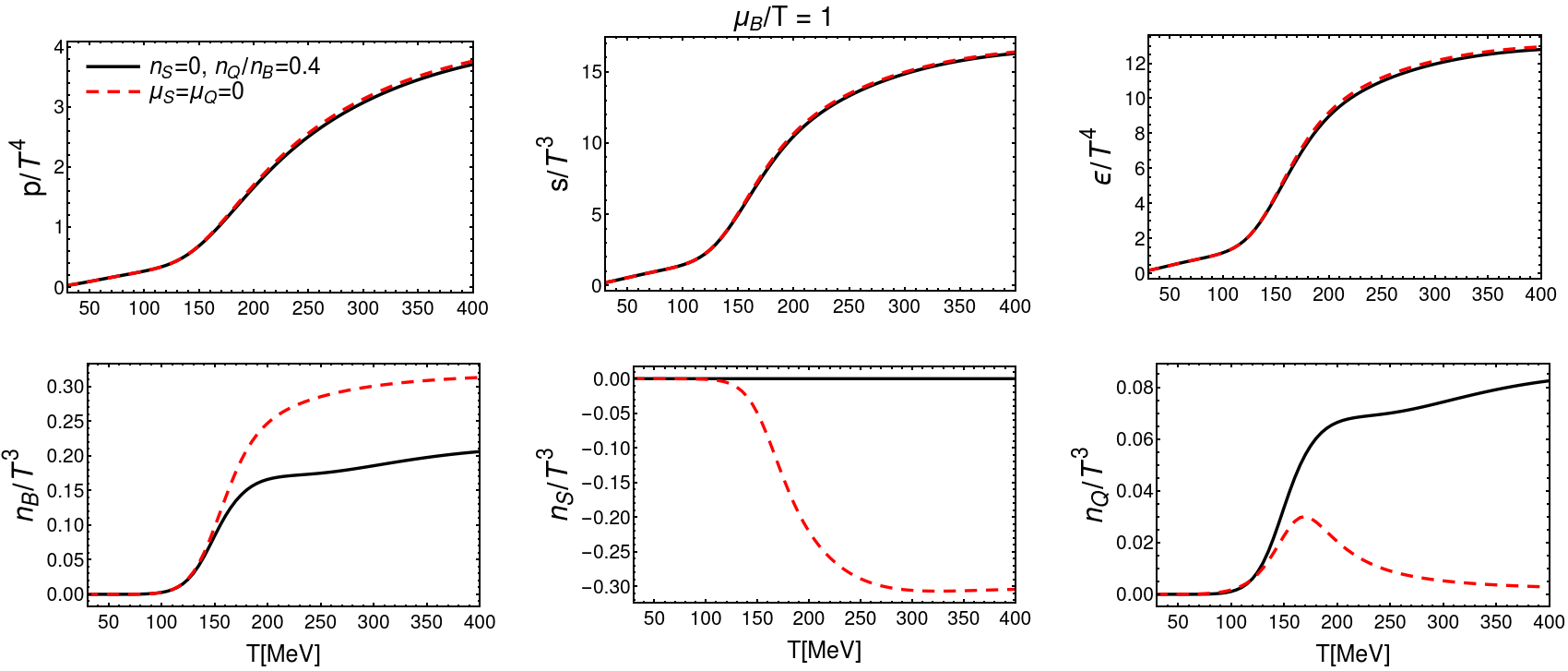} 

\includegraphics[width=\linewidth]{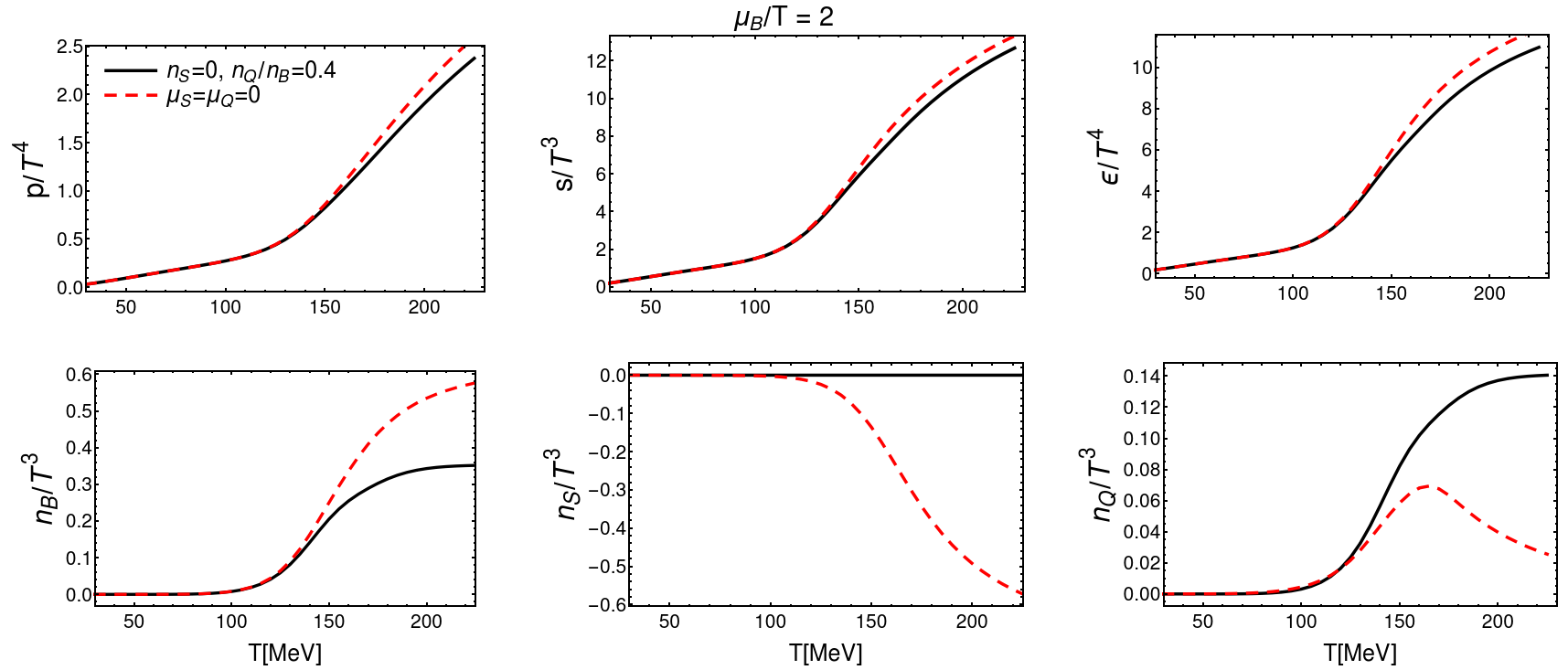} 
\caption{Normalized pressure, entropy density, energy density, baryonic, strangeness and electric charge densities are shown as functions of the temperature along the $\mu_B/T = 0.5$ (top panel), $\mu_B/T = 1.0$ (middle panel), $\mu_B/T = 2.0$ (bottom panel) lines. In all plots, the solid black curves indicate the case $\langle n_S\rangle=0$ and $\langle n_Q\rangle=0.4\langle n_B\rangle$, whereas the dashed red ones indicate the case $\mu_Q=\mu_S=0$.}
\label{fig4}
\end{figure*}

In Fig. \ref{fig4} we show the dependence of the normalized pressure, entropy density, energy density, baryonic, strangeness and electric charge densities on the temperature, along lines of constant $\mu_B/T=0.5,~1,~2$, both with $\langle n_S\rangle=0$, $\langle n_Q\rangle=0.4\langle n_B\rangle$ (solid black lines), and in the case $\mu_S = \mu_Q = 0$ (dashed red lines). We find that the thermodynamic quantities that are less sensitive to the chemical composition of the system do not show large discrepancies between the two scenarios, for all three values of $\mu_B/T$. On the other hand, when realistic conditions on the global chemical composition of the system are imposed, the baryon density is largely affected, and substantially decreased; the opposite effect is visible for the electric charge density, which is heavily enhanced.

\begin{figure}
\centering
\includegraphics[width=0.95\linewidth]{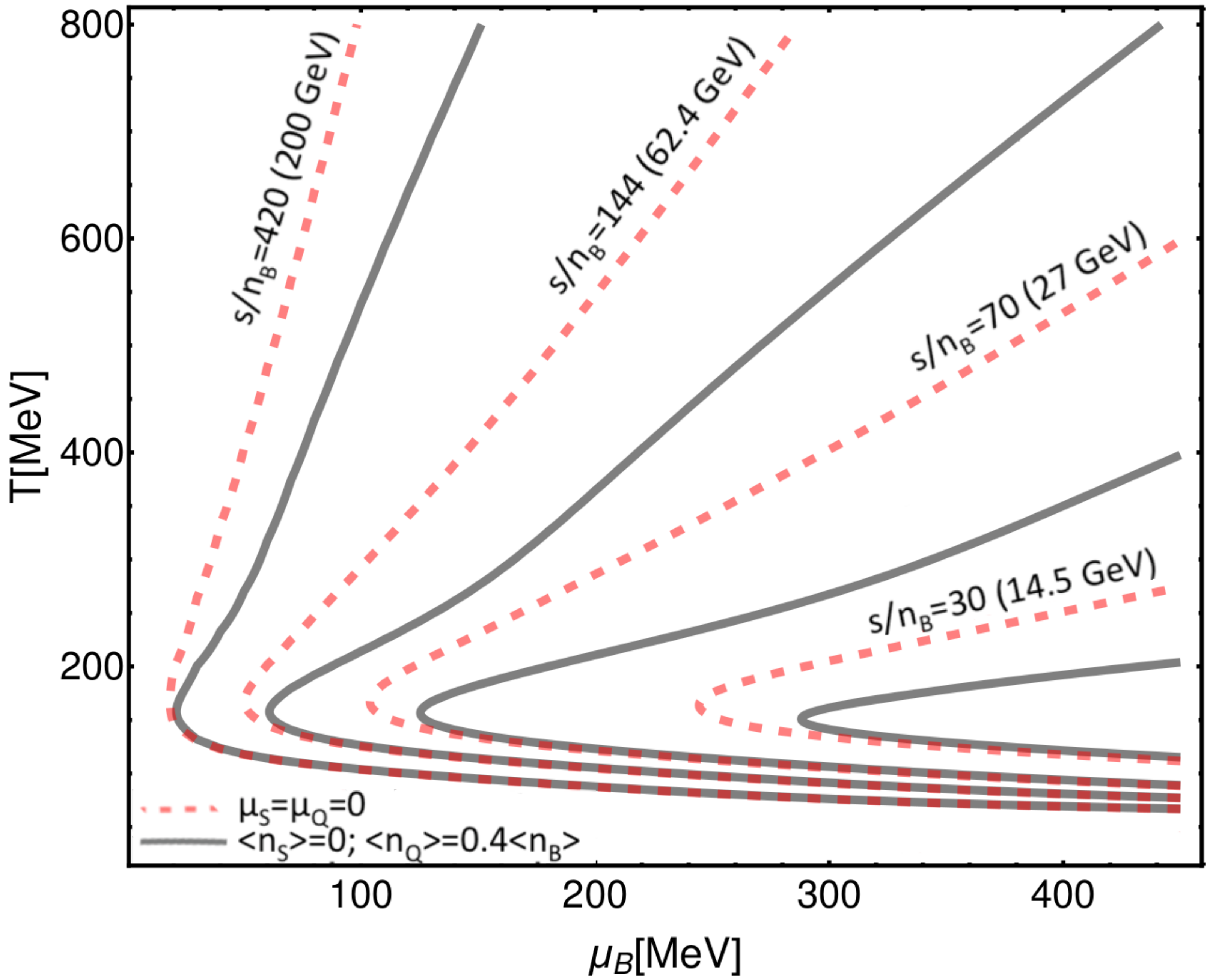} 
\caption{Isentropic trajectories in the $(T,~\mu_B)$ plane, for $s/n_B=420,~144,~70,~30$, corresponding to collision energies $\sqrt{s_{NN}}=200,~62.4,~27,~14.5$ GeV respectively. The solid black lines correspond to $\langle n_S\rangle=0$, $\langle n_Q\rangle=0.4\langle n_B\rangle$ while the dashed red lines to $\mu_S = \mu_Q = 0$.}
\label{fig6}
\end{figure}

\begin{figure}[h]
\centering
\includegraphics[width=0.95\linewidth]{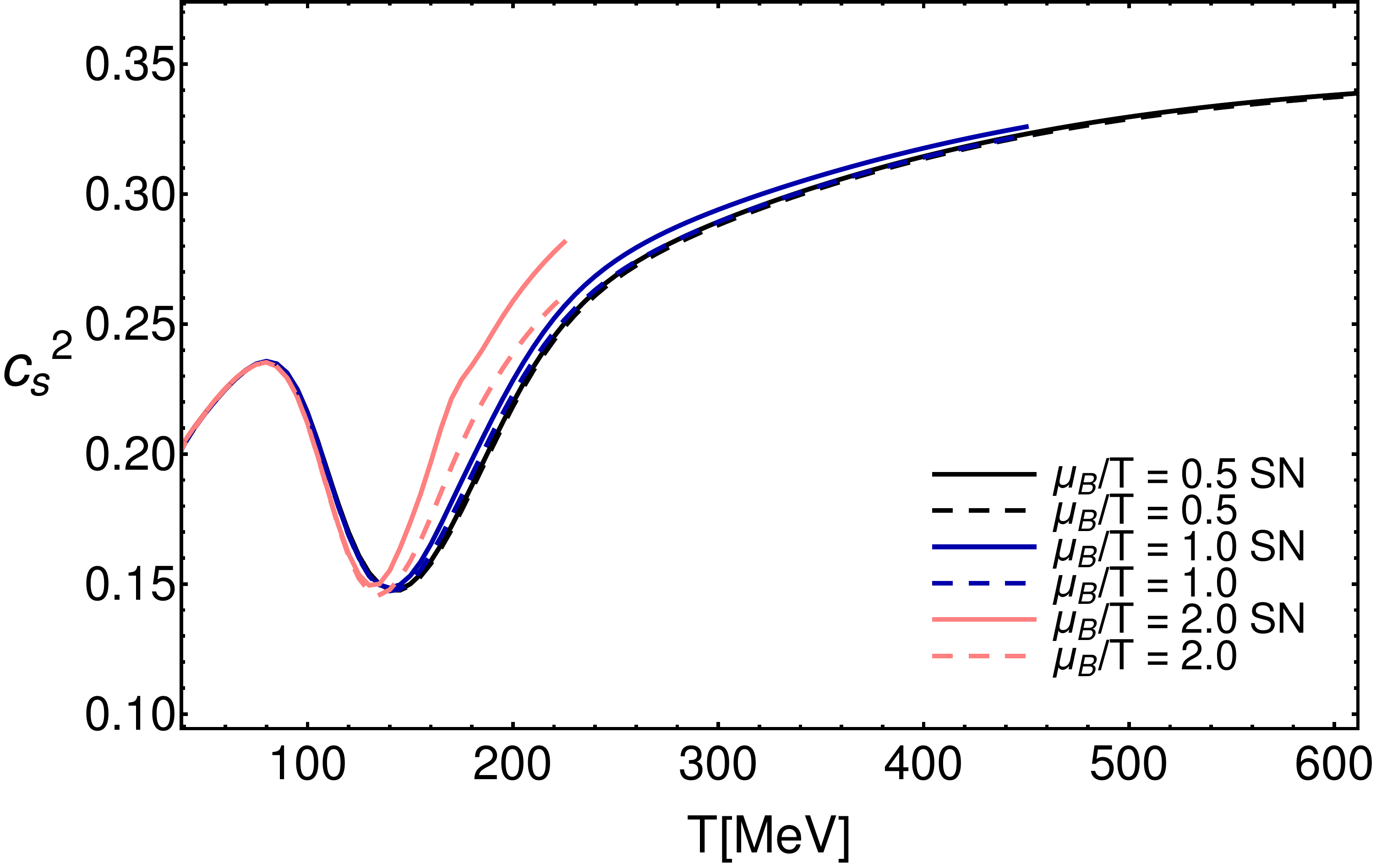} \\ \vspace{5mm}
\includegraphics[width=0.95\linewidth]{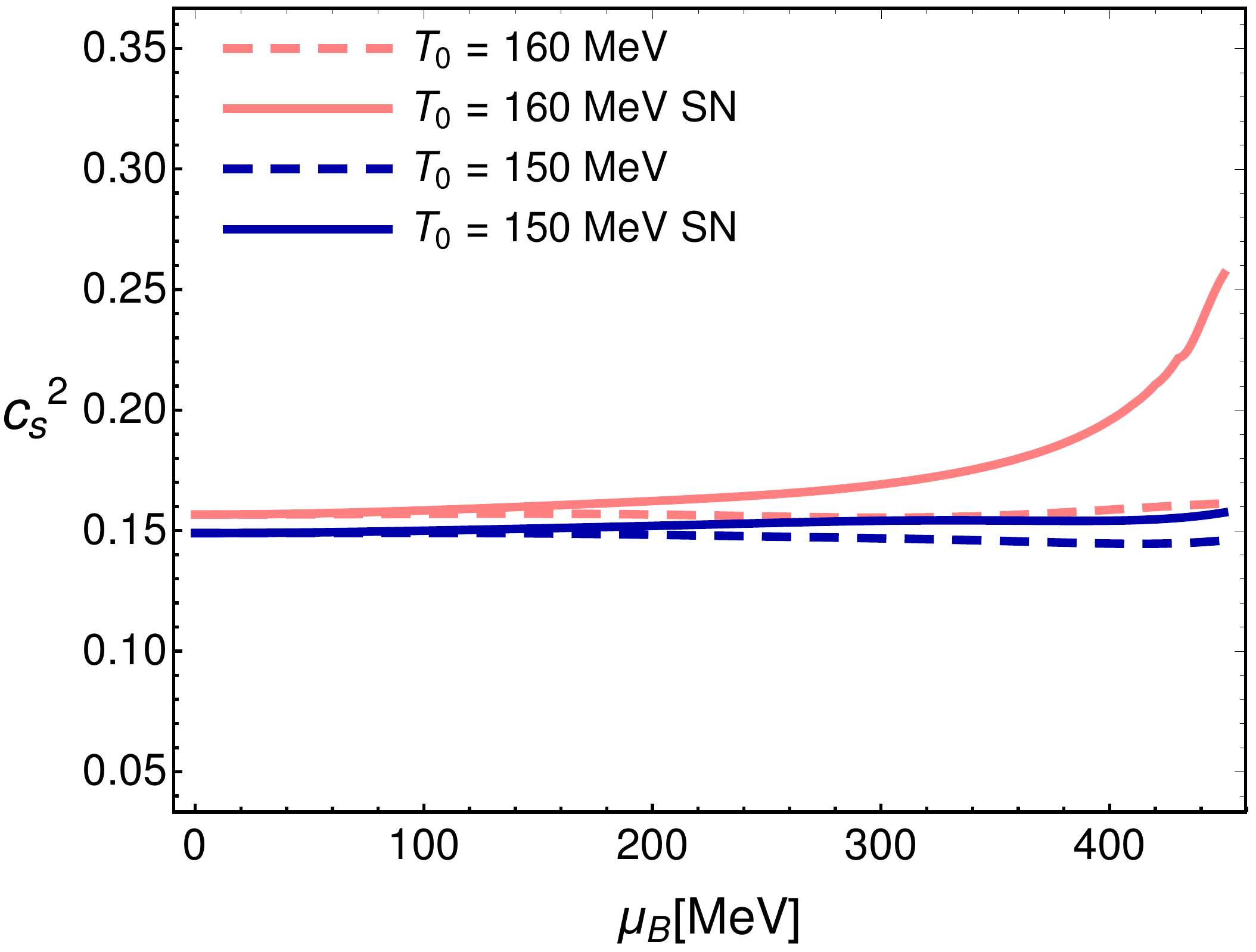}
\caption{(Upper panel) Temperature dependence of the speed of sound along lines of constant $\mu_B/T$. The solid lines correspond to $\langle n_S\rangle=0$, $\langle n_Q\rangle=0.4\langle n_B\rangle$ while the dashed ones to $\mu_S = \mu_Q = 0$. The curves for values of $\mu_B/T = 0.5,~1,~2$ are shown in black, blue/darker gray (this line stops at $T=450 \MeV$) and pink/lighter gray (this line stops at $T=225 \MeV$) respectively. (Lower panel) Behavior of the speed of sound along parametrized chemical freeze-out lines as in Eq. (\ref{eq:Cley_line}), with $T_{FO} (\mu_B=0) =  160 \MeV$ (pink/lighter gray lines) and $T_{FO} (\mu_B=0) =  150 \MeV$ (dark blue/darker gray lines). As in the upper panel, solid and dashed lines correspond to the cases with and without strangeness neutrality, respectively.}
\label{fig7}
\end{figure}

\begin{figure}[!h]
\centering
\includegraphics[width=0.95\linewidth]{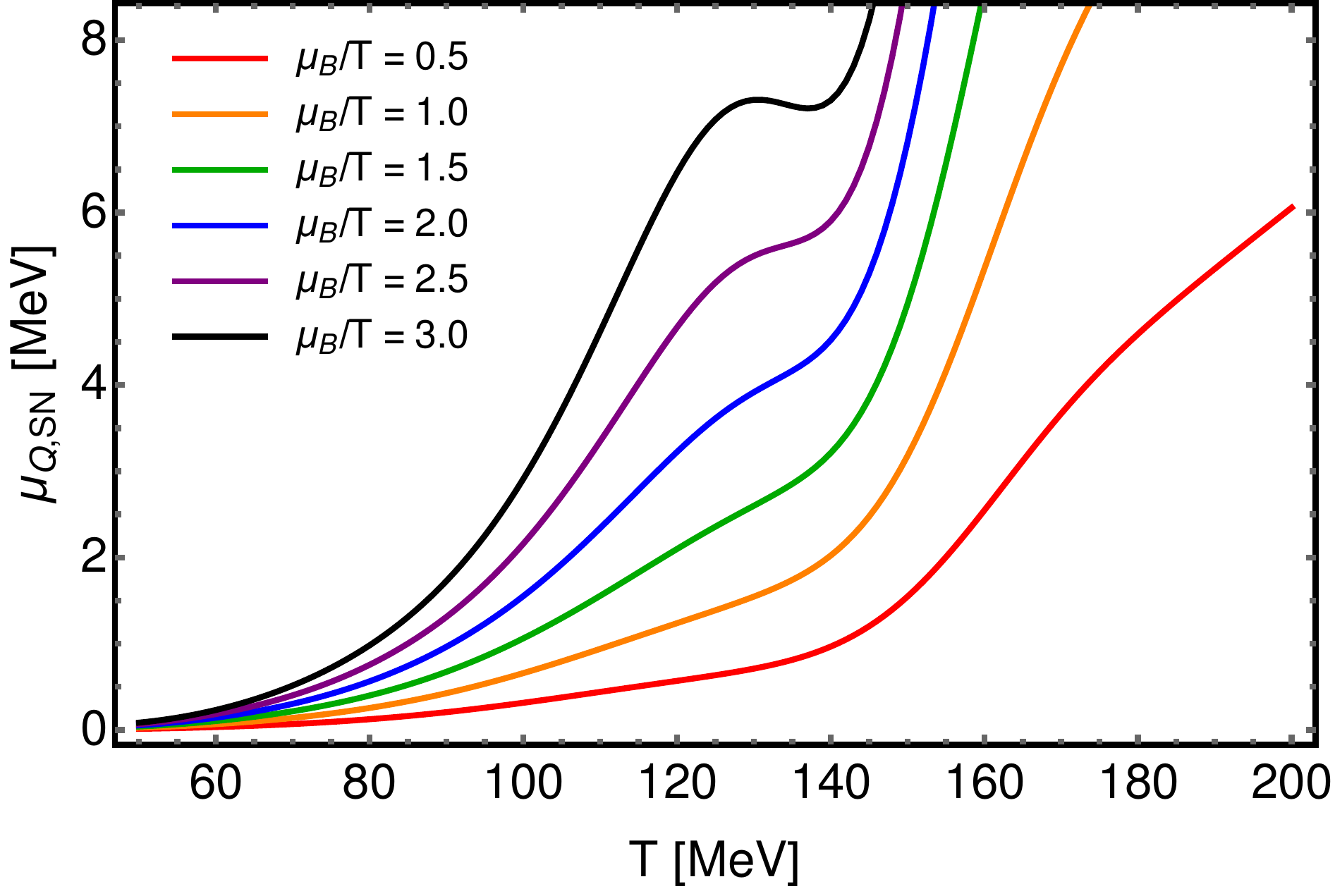} 
\caption{Temperature dependence of the electric chemical potential along lines of constant $\mu_B/T = 0.5 - 3$, in the case of strangeness neutrality.}
\label{fig8}
\end{figure}

Finally, we compare i) the isentropic trajectories, ii) the temperature dependence of the speed of sound along lines of constant $\mu_B/T$ and iii) the behavior of the speed of sound along parametrized chemical freeze-out lines between these two cases. The isentropic trajectories are shown in Fig. \ref{fig6} for selected values of $s/n_B$, which correspond to the indicated collision energies \cite{Gunther:2016vcp}. In the upper panel of Fig. \ref{fig7} we show the speed of sound as a function of the temperature along lines with $\mu_B/T=0.5,~1,~2$; the different colors correspond to different values of $\mu_B/T$. In the lower panel of Fig. \ref{fig7} we show the behavior of the speed of sound along two parametrized chemical freeze-out lines. These two freeze-out lines are shifted from the one presented in \cite{Cleymans:2005xv}, and have the form:
\begin{equation} \label{eq:Cley_line}
T_{FO}(\mu_B) = T_0 + b \mu_B^2 + c \mu_B^4 \, \, ,
\end{equation}
with $b = - 1.39 \cdot 10^{-4} \MeV^{-2}$ and $c = - 5.3 \cdot 10^{-11} \MeV^{-3}$; the two lines we show have $T_{FO} (\mu_B=0) =  160 \MeV$ and $T_{FO} (\mu_B=0) =  150 \MeV$. Both in Fig. \ref{fig6} and in Fig. \ref{fig7}, the solid lines correspond to $\langle n_S\rangle=0$, $\langle n_Q\rangle=0.4\langle n_B\rangle$ while the dashed lines to $\mu_S = \mu_Q = 0$. 

Since the EoS constructed in this work is a Taylor expansion carried out from lattice-QCD-calculated expansion coefficients, it is important to have an idea of the range of the validity of such expansion. It has been shown from lattice QCD simulations that the Taylor expansion of the Equation of State up to ${\cal O}(\mu_B^4)$ converges for $\mu_B/T \lesssim 2 - 2.5$ \cite{Bazavov:2017dus}, and the same can be said for our EoS. This roughly corresponds to a collision energy of $\sqrt{s} \gtrsim 10 \GeV$ \cite{Cleymans:2005xv}. In order to have a better idea of where a possible breakdown of its validity occurs, we show in Fig. \ref{fig8} the behavior of the electric chemical potential in the case with strangeness neutrality, along lines of constant $\mu_B/T = 0.5 - 3$. We see that a non-monotonic behavior appears around and above $\mu_B/T \sim 2.5$. This is in line with the expectation that the convergence of the Taylor series is guaranteed in the regime $\mu_B/T \lesssim 2.5$.  We note again that with the Taylor expansion approach used here, we do not expect to fully incorporate the constraints from imaginary $\mu_B$ -- and thus reproduce the Fourier harmonics from \cite{Vovchenko:2017xad} -- since for them the coverage of the region $\mu_B/T \leq \pi$ would be required. Applying the constraints from imaginary $\mu_B$ can be done in the near future to further improve our modeling of the QCD EoS, possibly concurrently with the inclusion of new continuum extrapolated lattice results.


\section{Conclusions}
In this manuscript, we constructed an equation of state for QCD at finite temperature and $B,~Q,~S$ chemical potentials, based on a Taylor series up to fourth power in the chemical potentials. Our methodology is based on a smooth merging between the HRG model and lattice QCD results for each one of the Taylor expansion coefficients; for all coefficients except $\chi_2^B$, the parameterization function is a ratio of up-to-ninth order polynomials. We provide all parameters in Tables \ref{table1},\ref{table2},\ref{table3}, so that our EoS can be readily used in the community. Furthermore, the code to generate the EoS and the tables for the thermodynamic quantities as functions of $T,~\mu_B,~\mu_S,\mu_Q$ is available at the link mentioned in Ref. \cite{code:2019}.

The Equation of State presented in this manuscript is important for the hydrodynamic description of the system created in heavy ion collisions at RHIC.  There are numerous  outstanding questions that remain to be understood at finite baryon densities that are influenced both by electric charge and strangeness. One recent surprise that arose from the first Beam Energy Scan was $\Lambda$ polarization, that indicates that the Quark Gluon Plasma may be the most vortical fluid known to humanity \cite{STAR:2017ckg}.  However, considering that $\Lambda$'s are simultaneously both strange particles and baryons, polarization studies should be done in hydrodynamic simulations that also consider all three conserved charges because of this interplay between strangeness and baryon number.  As previously mentioned, this $BQS$ equation of state can help shed light on the possible flavor hierarchy of freeze-out temperatures as well as the chiral magnetic effect.  A variety of dynamical observables of conserved charges (e.g. kaon flow harmonics) have already been measured at the Beam Energy Scan I and many others are planned for the Beam Energy Scan II, which may help to further constrain the location of a possible critical point.  

Finally, we point out that strange hadrons make up roughly $10\%$ of all measured  hadrons (assuming the kaon to pion ratio is a reasonable estimate for the ratio of all final state hadrons) and we can primarily only measure charged particles\footnote{Some neutral particles can be reconstructed from their daughter particles e.g. $\pi^0\rightarrow \gamma\gamma$}. Thus, a $BQS$ equation of state is required for a fully consistent description of the Quark Gluon Plasma at finite densities.  Relativistic hydrodynamics in the presence of multiple conserved charges obtains cross terms that affect the transport coefficients \cite{Denicol:2012vq,Denicol:2012cn,Greif:2017byw}.  Thus, it is misleading to extract transport coefficients at finite baryon densities only considering finite baryon number and not also finite strangeness and electric charge.  Furthermore, transport coefficients of different conserved charges have different characteristic temperatures, which further complicates the picture at large densities \cite{Rougemont:2017tlu}. The consequences are still under development, but it is certain that a $BQS$ equation of state is a vital first step to take into account any of these effects.

At this point, our reconstructed $BQS$ equation of state only consists of a cross-over transition.  Unlike a previous work where an equation of state at finite $\mu_B$ was coupled to the 3D Ising model in order to study criticality \cite{Parotto:2018pwx}, such an endeavor with three conserved charges would be significantly more complicated.  While the term ``critical point" is used, there might actually be a critical line or even critical plane once one considers the full three dimensional space of $\mu_B$, $\mu_S$, and $\mu_Q$.  Since there are large fluctuations in $T$, $\mu_B$, $\mu_S$, and $\mu_Q$ throughout the evolution of a single event  \cite{Becattini:2007ci,Shen:2017bsr,Brewer:2018abr}, certain elements of the fluid might pass through a critical region at an entirely different combination of $T$, $\mu_B$, $\mu_S$, and $\mu_Q$.
\vspace{.5cm}

\section*{Appendix}
We list the values of the parameters in Eq. (\ref{param}) for each Taylor expansion coefficient in Table \ref{table1}.
The Stefan-Boltzmann limit for the coefficients have the following values: 
\begin{align}
& & \frac{p(T,0,0,0)}{T^4} &= \frac{19\pi^2}{36}, & & & \\
\chi_2^B &= \frac{1}{3}, & \chi_2^Q &=\frac{2}{3}, & \chi_2^S &= 1, \nonumber\\
\chi_{11}^{BQ} &= 0, & \chi_{11}^{BS} &=-\frac13, & \chi_{11}^{QS} &= \frac13, \nonumber\\
\chi_4^B &= \frac{2}{9\pi^2}, & \chi_4^Q &=\frac{4}{3\pi^2},  & \chi_4^S &=\frac{6}{\pi^2}, \nonumber \\
\chi_{31}^{BQ} &= 0, & \chi_{31}^{BS} &= -\frac{2}{9\pi^2}, & \chi_{31}^{QS} &= \frac{2}{9\pi^2}, \nonumber \\
\chi_{13}^{BQ} &= \frac{4}{9\pi^2}, & \chi_{13}^{BS} &= -\frac{2}{\pi^2}, & \chi_{13}^{QS} &= \frac{2}{\pi^2}
\nonumber\\
\chi_{22}^{BQ} &= \frac{4}{9\pi^2}, & \chi_{22}^{BS} &= \frac{2}{3\pi^2}, & \chi_{22}^{QS} &= \frac{2}{3\pi^2}
\nonumber \\
\chi_{211}^{BQS} &= \frac{2}{9\pi^2}, & \chi_{121}^{BQS} &= -\frac{2}{9\pi^2}, & \chi_{112}^{BQS} &= -\frac{2}{3\pi^2}
\nonumber
\end{align}

\begin{center}
\begin{sidewaystable*} 
	\vspace{8cm}

    \begin{tabular}{|c|c|c|c|c|c|c|c|c|c|c|} 
        \hline 
        & $a_0$ & $a_1$ & $a_2$ & $a_3$ & $a_4$ & $a_5$ & $a_6$ & $a_7$ & $a_8$ & $a_9$   \\
        \hline 
        $\chi_0(T)$ & $7.53891$ & $-6.18858$ & $-5.37961$ & $7.0875$ & $-0.97797$ & $0.0302636$ & $-$ & $-$ & $-$ & $-$ \\
        \hline 
        $\chi_2^Q(T)$ & $-1.254$ & $13.7781$ & $-20.8361$ & $11.4637$ & $-1.52145$ & $0.0563044$ & $-$ & $-$ & $-$ & $-$ \\
        \hline 
        $\chi_2^S(T)$ & $0.728917$ & $-1.73212$ & $1.61219$ & $-0.706361$ & $0.192223$ & $-0.0164219$ & $-0.0040308$ & $0.00044212$ & $-$ & $-$  \\
        \hline 
        $\chi_{11}^{BQ}(T)$ &  $0.611997$ & $0.260951$ & $0.439882$ & $4.04624$ & $-0.492197$ & $-0.479177$ & $0.108023$ & $-0.000271088$ & $-$ & $-$  \\ 
        \hline 
        $\chi_{11}^{BS}(T)$ & $-3.42744$ & $0.0807472$ & $0.155933$ & $1.76331$ & $-0.350538$ & $-0.547143$ & $0.0641196$ & $-0.000271926$ & $-$ & $-$ \\
        \hline 
        $\chi_{11}^{QS}(T)$ & $0.975914$ & $-2.2118$ & $1.99441$ & $-0.710665$ & $0.100002$ & $-0.00437518$ & $-$ & $-$ & $-$ & $-$  \\       
        \hline 
        $\chi_4^B(T)$ & $0.0697892$ & $-0.0759267$ & $0.0270699$ & $-0.00183789$ & $-0.00102026$ & $0.000248834$ & $-0.0000205803$ & $5.78113\cdot10^{-7}$ & $-$ & $-$ \\
        \hline 
        $\chi_4^Q(T)$ & $0.519384$ & $-2.61484$ & $6.99796$ & $-9.37407$ & $5.50677$ & $-0.933273$ & $0.0628049$ & $-0.00149075$ & $-$ & $-$  \\
        \hline 
        $\chi_4^S(T)$ & $3.99178$ & $-10.8564$ & $11.4807$ & $-5.56961$ & $1.43254$ & $-0.204083$ & $0.0152834$ & $-0.00047076$ & $-$ & $-$ \\
        \hline 
        $\chi_{31}^{BQ}(T)$ & $0.000214078$ & $-0.00277202$ & $0.0107602$ & $-0.0189801$ & $0.0163346$ & $-0.00649086$ & $0.00102683$ & $-0.0000118454$ & $-$ & $-$ \\
        \hline 
        $\chi_{31}^{BS}(T)$ & $-0.606637$ & $0.940635$ & $-0.609091$ & $0.211817$ & $-0.0423212$ & $0.0048043$ & $-0.000283315$ & $6.59604\cdot10^{-6}$ & $-$ & $-$ \\
        \hline 
        $\chi_{31}^{QS}(T)$ & $1.39052$ & $-2.95215$ & $2.99901$ & $-1.3976$ & $0.337495$ & $-0.0441243$ & $0.0029685$ & $-0.0000804859$ & $-$ & $-$ \\
        \hline 
        $\chi_{13}^{BQ}(T)$ & $1.33817$ & $-0.36966$ & $-7.73766$ & $12.6268$ & $-7.54688$ & $2.27058$ & $-0.380023$ & $0.0357606$ & $-0.00175991$ & $0.0000349795$ \\
        \hline 
        $\chi_{13}^{BS}(T)$ & $-0.0853497$ & $0.09878$ & $-0.0477156$ & $0.0124373$ & $-0.00188339$ & $0.000165099$ & $-7.72499\cdot10^{-6}$ & $1.47927\cdot10^{-7}$ & $-$ & $-$ \\
        \hline 
        $\chi_{13}^{QS}(T)$ & $0.23137$ & $-0.607108$ & $0.574083$ & $-0.232842$ & $0.0476026$ & $-0.00514917$ & $0.000279883$ & $-5.97476\cdot10^{-6}$ & $-$ & $-$ \\
        \hline 
        $\chi_{22}^{BQ}(T)$ & $0.131897$ & $-0.151923$ & $0.0728375$ & $-0.0188047$ & $0.00281673$ & $-0.000244096$ & $0.0000112936$ & $-2.14344\cdot10^{-7}$ & $-$ & $-$ \\
        \hline 
        $\chi_{22}^{BS}(T)$ & $0.0481773$ & $-0.0633491$ & $0.034631$ & $-0.0101557$ & $0.00171648$ & $-0.000166247$ & $8.49007\cdot10^{-6}$ & $-1.75132\cdot10^{-7}$ & $-$ & $-$ \\
        \hline 
        $\chi_{22}^{QS}(T)$ & $1.03006$ & $-2.50946$ & $2.44698$ & $-1.00851$ & $0.207566$ & $-0.0225078$ & $0.00122422$ & $-0.000026132$ & $-$ & $-$ \\
        \hline 
        $\chi_{211}^{BQS}(T)$ & $0.146608$ & $-0.533936$ & $0.834892$ & $-0.645642$ & $0.260112$ & $-0.0540238$ & $0.00527256$ & $-0.000182156$ & $-$ & $-$ \\
        \hline 
        $\chi_{121}^{BQS}(T)$ & $-1.27191$ & $2.11351$ & $-1.4598$ & $0.538497$ & $-0.113414$ & $0.0134801$ & $-0.000826188$ & $0.0000197941$ & $-$ & $-$ \\
        \hline 
        $\chi_{112}^{BQS}(T)$ & $-2.61752$ & $4.37997$ & $-3.00258$ & $1.08514$ & $-0.221478$ & $0.0253053$ & $-0.00148533$ & $0.0000342702$ & $-$ & $-$ \\
        \hline 
	\end{tabular}
        \caption{Parameters $a_0 - a_9$ for the parametrization of the temperature dependence of all coefficients $\chi_{ijk}^{BQS} (T)$,  with the functional form shown in  Eq. (\ref{param}). The $``-"$ symbols in the table indicate that, for most of the coefficients, it is enough to consider a ratio of polynomials of order lower than seven.}
    \label{table1}
\end{sidewaystable*}
\end{center}

\begin{widetext}
\begin{center}
\begin{sidewaystable} \small

	\begin{tabular}{|c|c|c|c|c|c|c|c|c|c|c|c|} 
        \hline 
        &$b_0$ & $b_1$ & $b_2$ & $b_3$ & $b_4$ & $b_5$ & $b_6$ & $b_7$ &  $b_8$ & $b_9$ & $c_0$ \\
        \hline 
        $\chi_0(T)$ & $2.2453$ & $-6.02568$ & $15.3737$ & $-19.6331$ & $10.24$ & $0.799479$ & $-$ & $-$ & $-$ & $-$ & $-$ \\
        \hline
        $\chi_2^Q(T)$ & $-2.08695$ & $22.3712$ & $-33.4035$ & $19.9497$ & $-6.67937$ & $4.1127$ & $-$ & $-$ & $-$ & $-$ & $-$ \\ 
        \hline
        $\chi_2^S(T)$ & $0.634185$ & $-0.484646$ & $-3.02879$ & $7.29526$ & $-5.94029$ & $0.954829$ & $0.782178$ & $0.0848009$ & $-$ & $-$ & $0.00083$ \\ 
        \hline
        $\chi_{11}^{BQ}(T)$ & $506.969$ & $2.07112$ & $-1310.73$ & $-47.3907$ & $1855.62$ & $207.417$ & $-2635.03$ & $1616.16$ & $-$ & $-$ & $-$ \\ 
        \hline
        $\chi_{11}^{BS}(T)$ & $8.81578$ & $4.53879$ & $70.1272$ & $-212.977$ & $-287.925$ & $1688.03$ & $-2130.95$ & $901.004$ & $-$ & $-$ & $-$ \\ 
        \hline
        $\chi_{11}^{QS}(T)$ & $2.94254$ & $-5.97226$ & $4.37484$ & $0.723152$ & $-4.3139$ & $3.70245$ & $-$ & $-$ & $-$ & $-$ & $0.00012$ \\ 
        \hline
        $\chi_4^B(T)$ & $3.3139$ & $-2.34182$ & $-3.05239$ & $0.281088$ & $3.36387$ & $-1.47861$ & $0.232943$ & $-0.00920141$ & $-$ & $-$ & $-$ \\ 
        \hline
        $\chi_4^Q(T)$ & $2.78757$ & $-7.70015$ & $7.34828$ & $5.60254$ & $-16.5647$ & $10.1847$ & $-1.46422$ & $0.258243$ & $-$ & $-$ & $-$ \\ 
        \hline
        $\chi_4^S(T)$ & $7.26105$ & $-25.0961$ & $41.2002$ & $-31.1539$ & $-3.87268$ & $19.7369$ & $-9.31673$ & $2.02404$ & $-$ & $-$ & $-$ \\ 
        \hline
        $\chi_{31}^{BQ}(T)$ & $0.628355$ & $-1.27107$ & $-0.0555062$ & $0.801392$ & $0.649844$ & $-0.248501$ & $-1.16057$ & $0.662302$ & $-$ & $-$ & $-0.00007$ \\ 
        \hline
        $\chi_{31}^{BS}(T)$ & $22.8266$ & $-19.1507$ & $-33.6479$ & $25.4636$ & $17.3853$ & $-0.671223$ & $-19.7378$ & $9.96533$ & $-$ & $-$ & $-$ \\ 
        \hline
        $\chi_{31}^{QS}(T)$ & $52.129$ & $-92.6007$ & $24.1788$ & $32.9419$ & $-12.5404$ & $-1.67767$ & $1.02439$ & $0.502227$ & $-$ & $-$ & $-$ \\ 
        \hline
        $\chi_{13}^{BQ}(T)$ & $32.3922$ & $-36.2407$ & $-44.2609$ & $31.2543$ & $50.794$ & $17.5211$ & $-7.80941$ & $-13.3867$ & $-118.309$ & $93.7845$ & $-$ \\ 
        \hline
        $\chi_{13}^{BS}(T)$ & $0.285383$ & $0.769297$ & $-3.15803$ & $1.59797$ & $3.54785$ & $-0.652119$ & $-6.48277$ & $4.28691$ & $-$ & $-$ & $-$ \\ 
        \hline
        $\chi_{13}^{QS}(T)$ & $1.12154$ & $-2.86563$ & $2.35378$ & $-0.14257$ & $-0.827056$ & $0.35061$ & $-0.0544297$ & $0.125906$ & $-$ & $-$ & $-$ \\ 
        \hline
        $\chi_{22}^{BQ}(T)$ & $2.46229$ & $-1.78965$ & $3.86743$ & $-3.007$ & $-4.28013$ & $0.190242$ & $3.36159$ & $-0.215634$ & $-$ & $-$ & $-$ \\ 
        \hline
        $\chi_{22}^{BS}(T)$ & $0.505109$ & $0.555159$ & $-2.50987$ & $0.346874$ & $2.47285$ & $0.611415$ & $-3.84829$ & $2.02716$ & $-$ & $-$ & $-$ \\ 
        \hline
        $\chi_{22}^{QS}(T)$ & $15.1999$ & $-40.1845$ & $44.1416$ & $-19.6254$ & $-13.5991$ & $25.2683$ & $-12.6079$ & $2.72985$ & $-$ & $-$ & $-$ \\ 
        \hline
        $\chi_{211}^{BQS}(T)$ & $5.80204$ & $-15.5399$ & $5.25306$ & $18.444$ & $-1.81185$ & $-20.1787$ & $-4.61059$ & $13.9429$ & $-$ & $-$ & $-$ \\ 
        \hline
        $\chi_{121}^{BQS}(T)$ & $56.5761$ & $-106.452$ & $123.146$ & $-162.408$ & $94.5282$ & $51.273$ & $-77.7255$ & $29.3669$ & $-$ & $-$ & $-$ \\ 
        \hline
        $\chi_{112}^{BQS}(T)$ & $43.2755$ & $-108.526$ & $180.836$ & $-134.256$ & $-38.6051$ & $46.669$ & $6.94258$ & $17.7581$ & $-$ & $-$ & $-$ \\ 
        \hline 
    \end{tabular}
       \caption{Parameters $b_0 - b_9$ and $c_0$ for the parametrization of the temperature dependence of all coefficients $\chi_{ijk}^{BQS} (T)$,  with the functional form shown in  Eq. (\ref{param}). The $``-"$ symbols in the table indicate that, for most of the coefficients, it is enough to consider a ratio of polynomials of order lower than nine.}
    \label{table2}

	\vspace{16mm}

    \begin{tabular}{|c|c|c|c|c|c|} 
        \hline 
         & $h_1$ & $h_2$ & $f_3$ & $f_4$ & $f_5$    \\
        \hline 
        $\chi_2^B(T)$ & $-0.325372$ & $0.497729$ & $0.148987$ & $6.66388$ & $-5.07725$ \\
        \hline 
    \end{tabular}    
    
    \caption{Parameters for the parametrization of the temperature dependence of all coefficients $\chi_{ijk}^{BQS} (T)$,  with the functional form shown in  Eq. (\ref{param}). The $``-"$ symbols in the table indicate that, for most of the coefficients, it is enough to consider a ratio of polynomials of order lower than seven.}
    \label{table3}
\end{sidewaystable}
\end{center}
\end{widetext}

\section*{Acknowledgements}
This  material  is  based  upon  work  supported  by  the National  Science  Foundation  under  grants  no. PHY-1654219 and OAC-1531814 and by the U.S. Department of  Energy,  Office  of  Science,  Office  of  Nuclear  Physics, within the framework of the Beam Energy Scan Theory (BEST) Topical Collaboration.  We also acknowledge the support from the Center of Advanced Computing and Data Systems at the University of Houston. J.N.H. acknowledges support from the US-DOE Nuclear Science Grant No. DE-SC0019175.

\bibliography{biblio2}

\end{document}